\documentclass[12pt]{article}
\usepackage{graphicx}
\usepackage{epsfig}
\usepackage[figuresright]{rotating}
\usepackage{amsmath}
\usepackage{amssymb}
\usepackage{amsfonts}
\usepackage{array}

\renewcommand{\thefootnote}{\fnsymbol{footnote}}

\usepackage{cite}
\usepackage{amssymb}
\usepackage{amsfonts}

\begin{document}

\begin{center}
{\LARGE{\bf Dispersion theory of nucleon polarizabilities
 and  outlook on chiral effective field theory\footnote{Contribution
prepared for the  workshop ``Compton scattering off Protons and Light Nuclei
pinning down the nucleon polarizabilities'' ETC* Trento Italy, July 29 -
 Aug. 2, 2013} 
}}\\
[1ex]
Martin Schumacher\footnote{E-mail:
 mschuma3@gwdg.de}\\ 
II.  Physikalisches Institut der Universit\"at G\"ottingen,
Friedrich-Hund-Platz 1\\ D-37077 G\"ottingen, Germany
\end{center}

\renewcommand{\thefootnote}{\arabic{footnote}}
\setcounter{footnote}{0}

\begin{abstract}
Based on Compton scattering and meson photoproduction data
the polarizabilities of the nucleon 
are precisely studied and well understood due to  recent experimental 
and theoretical work using  nonsubtracted dispersion relations. The {\it
  recommended}  experimental values  are  
$\alpha_p=12.0\pm 0.6$, $(12.0)$,
$\beta_p= 1.9\mp 0.6$, $(1.9)$,  
$\alpha_n=12.5 \pm 1.7$, $(12.7\pm 0.9)$,
$\beta_n= 2.7 \mp 1.8$, $(2.5\mp 0.9)$
in units of $10^{-4}$fm$^3$ and
$\gamma^{(p)}_\pi=-36.4\pm 1.5$, $(-36.6)$, 
$\gamma^{(n)}_\pi= +58.6 \pm 4.0$, $(58.3)$,
$(\gamma^{(p)}_0=-0.58\pm 0.20)$, $(\gamma^{(n)}_0=0.38\pm 0.22)$
in units of 
$10^{-4}$fm$^4$ \cite{schumacher13}. The numbers given in parentheses 
are predicted values. It is shown that all versions of chiral
effective field theories
applied in analyses of nucleon polarizabilities and Compton scattering
ignore essential effects of
$\omega$, $\rho$ and $\sigma$ exchanges and of pseudoscalar $\pi$N
coupling. 
\end{abstract}

\section{Introduction}

The description of Compton scattering by the nucleon via dispersion theory
was developed at the beginning  of the 1960s \cite{hearn62}. The theory
made use of relations which  may be denoted 
as $s$-channel dispersion
relations and $t$-channel dispersion relations. The singularities entering
into the $s$-channel dispersion relations may be taken from meson
photoproduction experiments, whereas the singularities entering into
the $t$-channel dispersion relations are related to $\pi\pi$ pairs created in 
two-photon reactions in case of the scalar $t$-channel, and to the $\pi^0$
meson in case of the pseudoscalar $t$-channel. The first application 
of the scalar $t$-channel to the polarizabilities of the nucleon
came with the 
work of J. Bernabeu, T.E.O Ericson et al. (BEFT) in 1974 \cite{bernabeu74} 
where it was shown 
that the largest part of the electric polarizability and the total
diamagnetic polarizability are due to the $t$-channel. The smaller part
of the electric polarizability is due to the ``pion cloud'' showing up
as a nonresonant meson photoproduction process. The paramagnetic
polarizability is mainly due to the photoabsorption cross section
provided by the $P_{33}(1232)$ resonance. 

A convenient version of a dispersion theory applicable in a wide
angular interval and at energies up to 1 GeV was developed by L'vov et al.
\cite{lvov97}.
This dispersion theory is of the fixed-$t$ variety where $s$-channel
integrals are carried out along integration paths at constant $t$,
and the $t$-channel contributions are  taken care of in the form of
``asymptotic'' contributions, being an equivalent of  the $t$-channel
contributions. In principle this version of dispersion
theory is completely equivalent to other versions as there are fixed-$\theta$
dispersions theories or hyperbolic dispersion theories. At  first 
sight there appears to be a disadvantage in case of fixed-$t$ dispersion
theories because the integration paths are partly located in the
unphysical range of the scattering plane. However, it has been shown by L'vov
et al. \cite{lvov97}
that this only leads to minor technical  problems which can be solved without
loss of 
precision. The validity of this latter statement has been clearly demonstrated
by experiments, showing that fixed-$t$ dispersion theories 
lead to a precise representation of the experimental differential cross
sections  in the
whole angular interval and at energies up to 1 GeV
\cite{lvov97,galler01,wolf01}. In this respect the nonsubtracted dispersion
theory differs from the subtracted dispersion theory \cite{drechsel03} which
looses validity already at the peak energy of the $\Delta$ resonance.
For the spin-independent
$t$-channel
contribution the assumption was made that it is possible to represent it via
a scalar $\sigma$-meson pole-term, in analogy to the well-known 
pseudoscalar $\pi^0$ pole
term as entering into spindependent amplitudes. With this representation 
it was possible to arrive at agreement with experimental data 
in the angular and energy ranges described above, whereas without this 
representation of the pole term there remained a large discrepancy between
prediction and the experimental data.
This implies
that at energies of the second resonance region of the nucleon and
large scattering angles the $\sigma$-meson pole makes a dominant
contribution  to the Compton differential cross section. Furthermore,
the $\sigma$-meson mass was determined to be $m_\sigma=600$ MeV 
\cite{galler01,wolf01} in agreement
with  other available data.
These first investigations published in 2001 \cite{galler01,wolf01}
remained preliminary because 
the $\sigma$-meson pole was only an ansatz at that time 
and there was  an uncertainty about
its validity.    
This uncertainty has been  removed in later investigations where it was shown
that the $\sigma$-meson pole term has a very firm theoretical basis
and that the quantitative predictions calculated  from  well 
known properties of the $\sigma$-meson are  precise
(see \cite{schumacher13} for a
summary). 
As a conclusion we may state that the
dispersion theory of L'vov et al. \cite{lvov97}
is precise and well tested. Therefore, there
are good reasons to trust in the evaluations of 
electromagnetic polarizabilities
from low-energy Compton-scattering data where use is made of  this type of
dispersion theory. This is the case for all experimental data which are
summarized in \cite{schumacher05,schumacher13} leading to the 
{\it  recommended} nucleon polarizabilities given in the abstract. 
For the neutron also electromagnetic scattering of slow neutrons
has been taken into account.
Furthermore, 
these {\it recommended} values are in excellent agreement with independent
predictions
obtained from high-precision CGLN amplitudes \cite{drechsel07}
and well-investigated
properties of the $\sigma$ meson without making use of experimental
Compton differential
cross sections \cite{schumacher13}.

In addition to dispersion theory $\chi$EFT plays a 
prominent 
r\^ole in current investigations of nucleon Compton scattering and
polarizabilities.
The present investigation is motivated by the fact that recently $\chi$EFT
 has been used as a tool for  analyses of
experimental 
differential cross sections for Compton scattering by the proton, and results
have been obtained which are considerable different from the well-founded
standard values $\alpha_p=12.0\pm 0.6$ and 
$\beta_p=1.9\mp 0.6$. 
Examples are the ChPT-investigation of Beane et al. \cite{beane03}
where the values $\alpha_p=12.1\pm 1.1\pm 0.5$, $\beta_p=3.4\pm 1.1 \pm 0.1$
have been obtained, and McGovern et al. \cite{mcgovern13} where 
$\alpha_p=10.65\pm 0.35 \pm 0.36$,
$\beta_p=3.15\pm 0.35\pm 0.36$ have been obtained.
These values have been included in the data listing of the
Particle Data Group \cite{PDG} though the magnetic polarizabilities $\beta_p$
of \cite{beane03,mcgovern13} deviate from the respective {\it recommended} 
value \cite{schumacher13} by a factor $\sim 1.7$
and the electric polarizability $\alpha_p$ of \cite{mcgovern13} 
by $\sim 1.7$ standard deviations. These new results are based on the same
 set of experimental data as the previous ones \cite{schumacher13} 
so that the deviations are solely due to differences in the 
methods of data analysis. It is obvious that the new analyses are
justified only if  they are superior to or at least of the same quality as  
the previous ones. The purpose of the
present work is to explore whether or not this is the case.

\section{Summary of results of dispersion theory}

In a recent article \cite{schumacher13} a complete description of the present
status of  
dispersion theory of nucleon Compton scattering and polarizabilities has been
given. 
The {\it recommended} experimental polarizabilities introduced in the 2005 
summary \cite{schumacher05} have been confirmed in the 2013 summary
\cite{schumacher13} except for a slight correction applied to
$\gamma^{(p)}_\pi$. The reason for this slight correction was due to the fact
that new high-precision analyses of CGLN amplitudes have become available
\cite{drechsel07} which made this  revision necessary.   These 
new high-precision analyses \cite{drechsel07} were also of great importance
in connection with 
the prediction of the $s$-channel components of the nucleon polarizabilities 
for all resonant and nonresonant excitation processes of the nucleon. 
Results of these analyses which are of interest here 
are summarized in Table \ref{tab1}.
\begin{table}[h]
\caption{$s$-channel electromagnetic polarizabilities compared with
  experimental data.}
\begin{center}
\begin{tabular}{|l|cc|cc|}
\hline
&$\alpha_p$& $\beta_p$& $\alpha_n$& $\beta_n$\\
\hline
$s$-channel prediction& $+4.48$ & $+9.44$ & $+5.12$& $+10.07$\\ 
experimental data&$12.0\pm 0.6$& $1.9\mp 0.6$ & $12.5\pm 1.7$ & $2.7\mp 1.8$\\
\hline
difference line 3 - line 2&$+7.5$&$-7.5$ & $7.4$ & $-7.4$ \\
\hline
\end{tabular}
\end{center}
\label{tab1}
\end{table}
As documented in line 4 of Table \ref{tab1} the $s$-channel predictions 
show a large deviation from
the experimental data and  these deviations are the same for the proton
and the neutron. Furthermore, the differences obtained for the electric and 
the magnetic polarizabilities given in line 4  only differ by  the signs. This
means that 
these differences cancel in case of  $\alpha+\beta$, i.e. $
(\alpha+\beta)^t\equiv 0$,  but make a dominant
contribution, {\it viz.} $(\alpha-\beta)^t$,   in case of $\alpha-\beta$.

As has been pointed out in \cite{schumacher13} the $t$-channel contribution
differs from the $s$-channel contribution due to the fact that the former
 can be
traced back to the mesonic structure of  the constituent quarks. This 
means that the  short-distance contribution \cite{lvov01} introduced in 
some versions of $\chi$EFT
may tentatively
also be viewed in terms of properties of the constituent quarks. 
In spite of this interesting similarity there is an essential 
difference between 
$\chi$EFT and dispersion theory due to the fact that dispersion theory provides
a method for a quantitative prediction of the $t$-channel contribution, 
whereas the $\chi$EFT prediction for the short-distance contribution
is treated as an adjustment, filling the gap between predictions and
experimental data.  This is the reason for naming them counterterms (c.t.)
with $\delta\alpha$ corresponding to the electric part and $\delta\beta$
corresponding to the magnetic part.

\subsection{Quantitative prediction of the $t$-channel component}

The $t$-channel component of the electromagnetic polarizabilities  
of the nucleon has been
first described by J. Bernabeo, T.E.O. Ericson et al (BEFT) \cite{bernabeu74}
in the following way: 
If we restrict ourselves in the calculation
of the $t$-channel absorptive part to intermediate states with two
pions with angular momentum $J\leq 2$, the sum rule 
takes the convenient form for calculations \cite{bernabeu74}:
\begin{eqnarray}
(\alpha-\beta)^t&=& 
 \frac{1}{16 \pi^2}\int^\infty_{4 m^2_\pi}\frac{dt}{t^2}\frac{16}{4m^2-t}
\left(\frac{t-4m^2_\pi}{t}\right)^{1/2}\Big[f^0_+(t)
  F^{0*}_{0}(t)\nonumber\\
&&-\left(m^2-\frac{t}{4}\right)\left(\frac{t}{4}-m^2_\pi\right)
f^2_+(t) F^{2*}_{0}(t)\Big],\label{BackSR}
\end{eqnarray}
where
$f^{(0,2)}_+(t)$ and $F^{(0,2)}_0(t)$ are the partial-wave helicity
amplitudes of the processes $N\bar{N}\to \pi\pi$ and 
$\pi\pi\to \gamma\gamma$ with angular momentum $J=0$ and $2$,
respectively, and isospin $I=0$.  
Though being the first who published the BEFT sum rule in its presently
accepted form, Bernabeu and Tarrach \cite{bernabeu74}
were not aware of the appropriate amplitudes to calculate the BEFT sum rule
numerically. Later on evaluations of Eq. (\ref{BackSR}) also remained rather
uncertain until recently, when Drechsel et al. \cite{drechsel03} and Levchuk
\cite{levchuk04} carried out calculations with good precision.
\begin{table}[h]
\caption{Numerical evaluation of the BEFT  sum rule and of the equivalent
$\sigma$-meson pole.
}
\begin{center}
\begin{tabular}{|l|l|}
\hline
$(\alpha_p-\beta_p)^t$
&authors \& methods\\
\hline
$+16.5$&{Drechsel,Pasquini,
Vanderhaeghen,2003}\cite{drechsel03}, (Eq. \ref{BackSR})\\
$+14.0$&Levchuk, 2004 \cite{levchuk04}, (Eq. \ref{BackSR})\\
$+15.2$&prediction based on the $\sigma$-meson pole 
\cite{schumacher13}, (Eq. \ref{pol8})\\
\hline
\end{tabular}
\end{center}
 \label{BEFtable}
\end{table}
The results obtained in this way are listed in lines 2 and 3 of Table
\ref{BEFtable}.
In the $t$-channel notation the two-photon process described  
by Eq. (\ref{BackSR}) may be written in the form
\begin{equation}
\gamma\gamma \to \sigma \to \pi\pi \to \sigma \to N\bar{N},
\label{pipiNN}
\end{equation}
i.e. by a pion pair in the intermediate state, coupled to two photons
on the one hand and to a $N\bar{N}$ pair on the other, via correlations
which may be understood as  $\sigma$ mesons. As has been justified in detail
in \cite{schumacher13} this composite intermediate state can be replaced by
\begin{equation}
\gamma\gamma\to \sigma \to N\bar{N}
\label{sigmaNN}
\end{equation}
which describes  the $t$-channel pole contribution in complete analogy to the
well known $\pi^0$ pole contribution
 \begin{equation}
\gamma\gamma\to \pi^0 \to N\bar{N}.  
\label{pinullNN}
\end{equation}
This leads to the prediction 
derived in  \cite{schumacher07b,schumacher09} 
\begin{equation}
(\alpha-\beta)^t=\frac{g_{\sigma NN}{\cal M}(\sigma\to\gamma\gamma)}{2\pi m^2_\sigma}
+\frac{g_{f_0 NN}{\cal M}(f_0 \to\gamma\gamma)}{2\pi m^2_{f_0}}
+\frac{g_{a_0 NN}{\cal M}(a_0\to\gamma\gamma)}{2\pi m^2_{a_0}}\tau_3,
\label{pol8}
\end{equation}
where the $\sigma$-meson part is given by
\begin{equation}
(\alpha-\beta)^t_{\sigma}=\frac{5
  \alpha_{em} g_{\pi NN}}{6\pi^2 m^2_\sigma f_\pi}= 15.2
\label{tsigma}
\end{equation}
with $\alpha_{em}=1/137.04$, 
$g_{\sigma NN}\equiv g_{\pi NN}=13.169\pm 0.057$,   
$f_\pi=(92.42\pm0.26)$ MeV, $m_\sigma=666$ MeV,
as derived in \cite{schumacher06} and 
given in line 4 of Table  \ref{BEFtable}. 
Now with $(\alpha+\beta)^t=0$ we arrive at
\begin{eqnarray}
&&\alpha^t=+\frac12 (\alpha-\beta)^t=+ 7.6\,\,\, (\sim 62\% \,\,\, {\rm of}
\,\,\,  \alpha),
\label{alphat}\\
&&\beta^t=-\frac12 (\alpha-\beta)^t=-7.6\,\,\, (100\% \,\,\, {\rm of}
\,\,\, \beta_{\rm dia})     \label{betat}
\end{eqnarray}
in excellent agreement with the findings in Table \ref{tab1}. It should 
be noted that the largest part of the electric polarizability $\alpha$
(63\% for the proton and 61\% for the neutron) and 100\% of the diamagnetic
polarizability $\beta_{\rm dia}$ are due to the $t$-channel.

The contributions
from the  $f_0(980)$ and $a_0(980)$ scalar meson entering into
Eq. (\ref{pol8})
are discussed in detail in \cite{schumacher13}. These contributions are small
and may be represented by 
\begin{equation}
\alpha(f_0(980),a_0(980)) =\pm (+0.3-0.4\, \tau_3).
\label{alphafa}
\end{equation}
The double,  $\pm$,  sign on r.h.s. of Eq. (\ref{alphafa}) indicates
that we leave 
the sign of $\alpha(f_0(980),a_0(980))$ undetermined and, as a consequence, 
include this quantity into the error when calculating the neutron
electromagnetic polarizability $\alpha_n$ from the experimental proton
electric polarizability $\alpha_p$ and the predicted difference 
$(\alpha_n - \alpha_p)$, leading to \cite{schumacher13}
\begin{equation}
\alpha_n=12.7\pm 0.9, \quad \beta_n=2.5\mp 0.9.
\label{aln}
\end{equation}
The errors given in (\ref{aln}) take into account the experimental error
of $\alpha_p$ and the error due to the $t$-channel contributions of the
$f_0(980)$ and $a_0(980)$ scalar mesons. The error of $\alpha_p$ is a measure
of the precision of the procedure in general and the error of the
$f_0(980)$ and $a_0(980)$ contributions a measure of an additional uncertainty
in case of the neutron.
This result for the neutron polarizabilities is extremely important because
it rests on very firm arguments for the $t$-channel component and 
on very precise experimental data for the 
CGLN amplitudes. We propose to use the prediction given in (\ref{aln})
as a benchmark for future high-precicion experiment on the neutron.

The conclusion we may draw from this result is that dispersion theory
provides us with a quantitative prediction of the three components of
the electromagnetic polarizabilities, as there are the $s$-channel nonresonant
and $s$-channel resonant excitations of the nucleon, and the $t$-channel
part which may be understood as scattering by the $\sigma$ meson while being
part of the structure of the constituent quark. This latter process is
expected to take place because the $\sigma$ meson mediates the generation of
mass of the constituent  quark via chiral symmetry breaking and, therefore,
has to be a part of the constituent-quark structure. This very consistent
result obtained from dispersion theory contrasts with the unspecified
short-distance contribution discussed in case of $\chi$EFT.

\subsection{Dependence of the polarizabilities on the photon energy}

\begin{figure}[h]
\centering\includegraphics[width=0.7\linewidth]{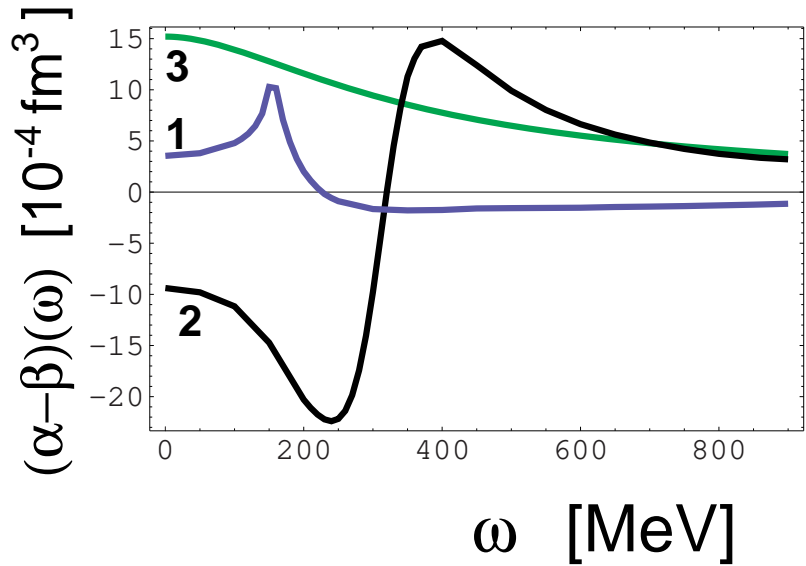}
\caption{Real part of the three  main  components of  $(\alpha-\beta)(\omega)$
for the proton. 1) Meson-cloud component as given by the $E_{0+}$ CGLN
amplitude. 2) Resonant component due to the $P_{33}(1232)$ nucleon
      resonance. 3) $t$-channel component as given by the $\sigma$-meson pole.
The  curve is calculated for the $\sigma$-meson mass
$m_\sigma=600$.
}
\label{albeomega-1}
\end{figure}

Compton scattering experiments aimed to determine the polarizabilities of 
the nucleon are carried out typically at energies above 50 MeV up to energies
well below the $\Delta$ peak. In the upper part of this energy interval 
fits to the experimental differential cross sections require a general
knowledge of the photon-energy,  
$\omega$, dependences of the polarizabilities. 
Dispersion relations evaluated in the backward direction at $\theta=\pi$
are  the appropriate tool to determine these $\omega$
dependences.  Figure \ref{albeomega-1} shows the result of the calculation
when including the empirical $E_{0+}$ CGLN amplitude, the $P_{33}(1232)$
resonance and the $\sigma$-meson pole contribution. The latter contribution
has been calculated for the mass $m_\sigma= 600$ MeV. 
This mass is the appropriate value because it fits the experimental
differential 
cross sections in the second resonance region of the proton. The dispersion
relation used for these calculations
may be found in \cite{schumacher13}.

\begin{figure}
\includegraphics[width=0.45\linewidth]{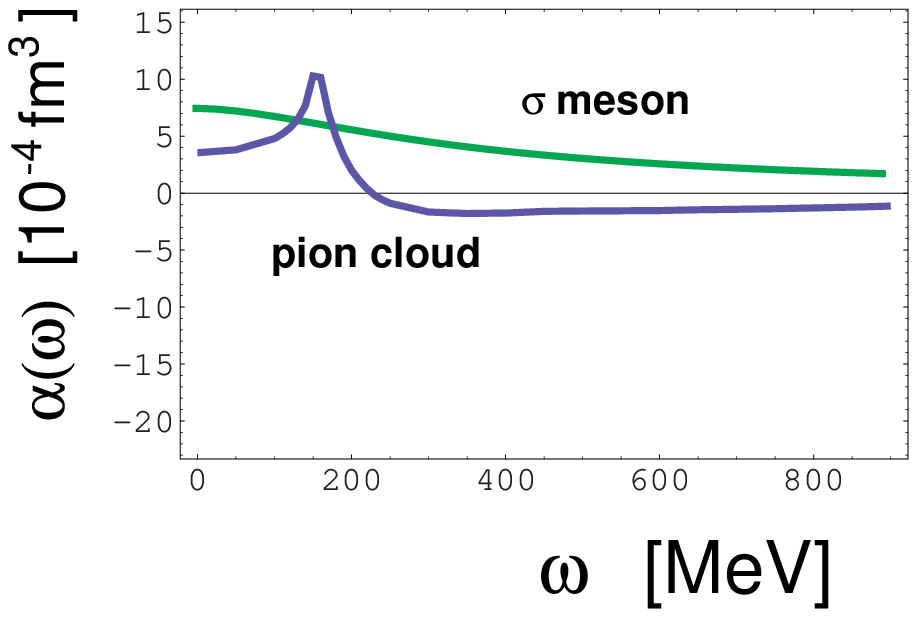}
\includegraphics[width=0.55\linewidth]{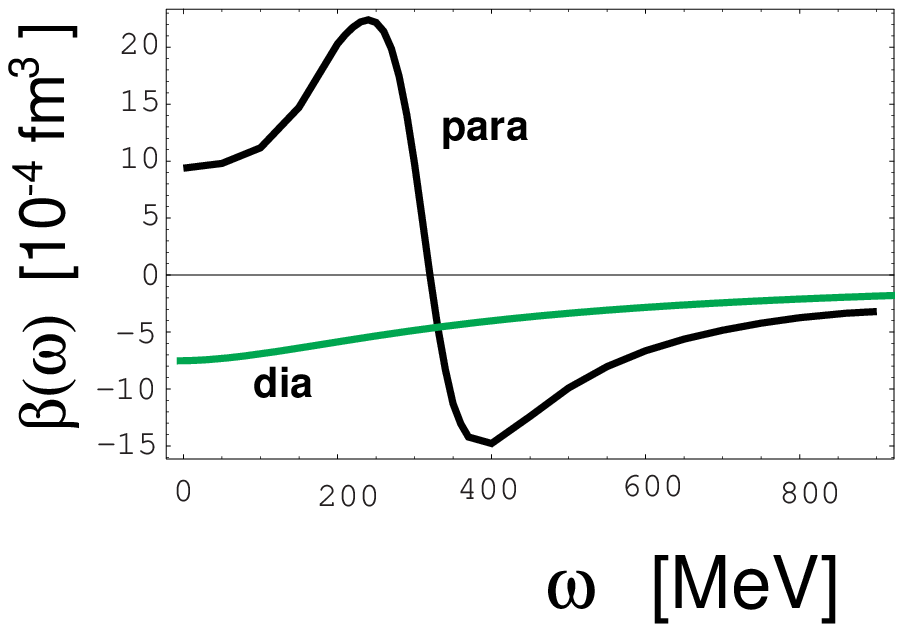}
\caption{Left panel: Energy dependent electric polarizability 
$\alpha(\omega)$ due to the 
pion cloud and the $\sigma$-meson pole (see Fig. \ref{albeomega-1}).
Right panel: Energy dependent magnetic polarizability $\beta(\omega)$
due to the $P_{33}(1232)$ resonance and
the $\sigma$-meson pole (see Fig. \ref{albeomega-1}).
}
\label{albeomega-2}
\end{figure}

It may be  of interest
to disentangle the curves shown in Figure \ref{albeomega-1} into an electric
part and a magnetic part. The results  obtained
are  shown in Figure  \ref{albeomega-2}.
 \begin{figure}
\includegraphics[width=0.45\linewidth]{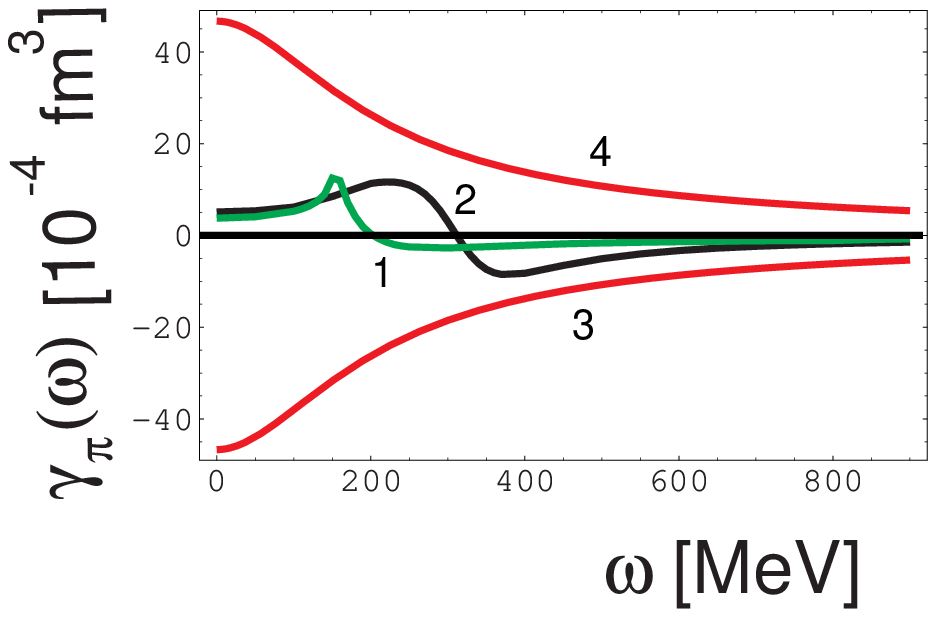}
\includegraphics[width=0.45\linewidth]{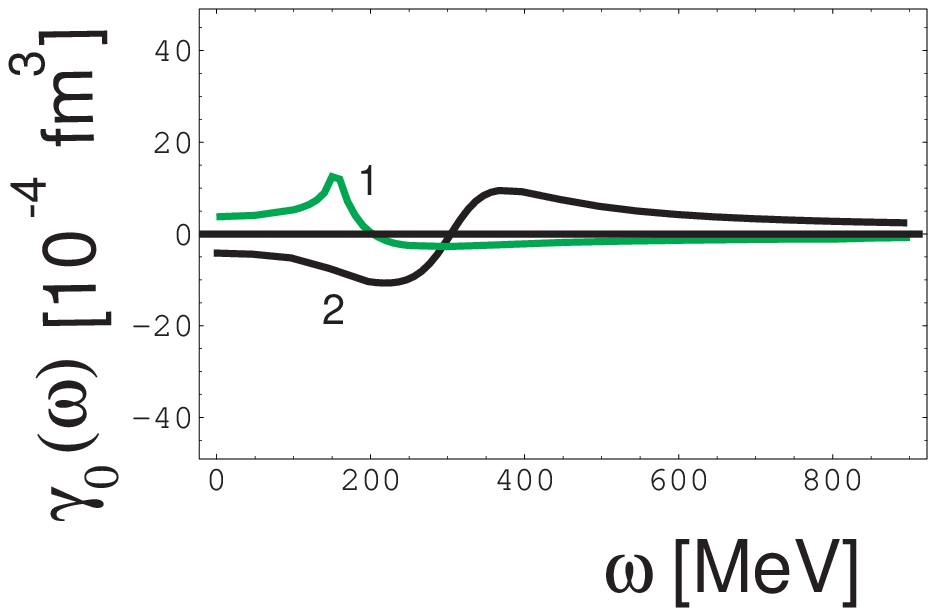}
\caption{Photon-energy dependent spinpolarizabilities. Left panel:
 Components of backward  spinpolarizabilities, 1) $E_{0+}$ component,
2) $P_{33}(1232)$ component, 3) $\pi_0$-pole component for the proton, 
4)    $\pi_0$-pole component for the neutron. Right panel: Components of
forward  spinpolarizabilities, 1) $E_{0+}$ component,
2) $P_{33}(1232)$ component}.
\label{gampi}
\end{figure}
The $\omega$-dependent electric polarizability as shown in the left
panel of Figure \ref{albeomega-2}
is a superposition of the $E_{0+}$  contribution
as given by the nonresonant photoabsorption cross section
and a positive part of  the $\sigma$-meson pole contribution.  In a similar way
the magnetic 
polarizability as shown in the right
panel of Figure \ref{albeomega-2}
can be traced back to the $P_{33}(1232)$ resonance as the main
component of the paramagnetic polarizability, and a negative part of
the $\sigma$-meson pole
contribution which represents the diamagnetic polarizability. 

For sake of completeness we also investigate the $\omega$ dependencies of the
spinpolarizabilities. In the left panel of Figure \ref{gampi} 
the components of the backward spinpolarizability are shown. There is a
constructive interference of the $E_{0+}$ component (curve 1) and the 
$P_{33}(1232)$ component (curve 2). The main components are due to the
$\pi^0$ $t$-channel  which lead to a destructive interference in case
of the proton and to a constructive interference in case of the neutron.
In case of the forward spinpolarizability as shown in the right panel
of Figure \ref{gampi} the $E_{0+}$ component (curve 1) 
and the $P_{33}(1232)$ component (curve 2) interfere destructively. 
This leads to very small values for the forward spinpolarizabilities
for the proton as well as for the neutron.
Furthermore, since the  $E_{0+}$ component is larger for the neutron than for
the proton, whereas the $P_{33}(1232)$ components are the same, it is not a
surprise that the forward spinpolarizabilities of the proton and neutron have 
different signs (see \cite{schumacher13} and references therein).

\section{Outlook on chiral perturbation theory}

Even though QCD is the correct theory for the strong interactions, it cannot
easily be used for computations at all energy and momentum scales. At energies
of the order of the nucleon mass the Nambu--Jona-Lasinio (NJL) model 
\cite{nambu61,vogl91,klevansky92,hatsuda94,bijnens96} works extremely well. 
In addition, the Linear $\sigma$ Model (L$\sigma$M)
\cite{gell-mann60,alfaro73} may be
applied where the aspect of spontanous symmetry breaking is exploited.
For practical applications it is convenient to make use of the Quark-Level
Linear $\sigma$ Model (QLL$\sigma$M) \cite{scadron13} which combines 
properties of the two models, NJL and L$\sigma$M. For the special problem
of nucleon polarizabilities   the QLL$\sigma$M of course cannot be applied to
the prediction of the polarizabilities in general, but it is very useful 
and extremely precise to make predictions on the basis of the QLL$\sigma$M
for the $t$-channel component \cite{schumacher13}. 
The $s$-channel component can only be reliably predicted  on the basis of
photomeson data by applying  dispersion relations. This has been described
in detail in the foregoing section.

Chiral perturbation theory dates back to a paper of Steven Weinberg 
\cite{weinberg79} which is concerned with phenomenological Lagrangians.
In this work, for simplicity, Weinberg ``integrates out'' whatever other
degrees of freedom may be present - nucleon, $\rho$ meson, $\sigma$ meson,
etc. -  and he considers only pions.  Furthermore, the derivative 
(pseudovector) coupling 
of the pion is introduced as a desired property  of the theory.

In case of Compton scattering by the nucleon Weinberg's concept is applied to
the $N\pi$-system \cite{bernard91,bernard92}. Later on the $\Delta$,
$\Delta\pi$ and short-distance degrees of freedom were also 
taken into account.

\subsection{Chiral perturbation theory and the ${\bf E_{0+}}$ CGLN amplitude}

\begin{figure}
\includegraphics[width=0.65\linewidth]{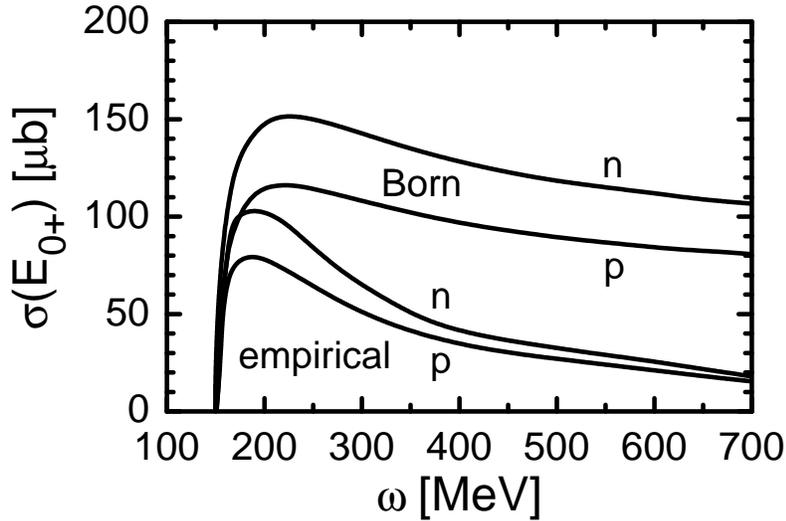}
\caption{Photoabsorption cross section $\sigma(E_{0+})$ due to $s$-wave 
single-pion photoproduction for the proton (p) and the neutron (n). The 
two upper curves have been calculated in the Born approximation. The two 
lower curves are
empirical results \cite{drechsel07,hanstein98,drechsel99}.}
\label{e0plus}
\end{figure}

In 1993 A. I. L'vov published a paper \cite{lvov93b} with the title 
``A dispersion look at
the chiral perturbation theory. Nucleon electromagnetic polarizabilities''.
In this paper it is shown  that chiral perturbation theory (ChPT) as discussed 
at that time \cite{bernard91,bernard92}
corresponds to dispersion theory applied to the photomeson 
$E_{0+}$ CGLN amplitude in the Born approximation. The results obtained
are 
\begin{equation}
\alpha^{\rm pion\,Born}_p = 7.3, \quad
\beta^{\rm pion\,Born}_p = -1.8,\quad
\alpha^{\rm pion\,Born}_n = 9.8, \quad
\beta^{\rm pion\,Born}_n = -0.9
\label{pionBorn}
\end{equation} 
in very good agreement with the prediction of ChPT \cite{bernard92}
as given in the second line of Table \ref{tab3}. By the same procedure 
the results of the heavy baryon version (HBChPT) have been reproduced 
by making use the shift $m_p\to \infty$ in appropriate parts of
the calculation. The numerical
results obtained in the HBChPT are given in the third line of Table
\ref{tab3}.
\begin{table}[h]
\caption{Predicted electromagnetic polarizabilities of the nucleon compared
  with  
experimental data. The predictions given in lines 2 and 3 correspond to
ChPT and HBChPT, respectively,  the prediction in
line 4 to the empirical  $E_{0+}$ CGLN amplitude. }
\begin{center}
\begin{tabular}{|l|cc|cc|}
\hline
&$\alpha_p$&$\beta_p$&$\alpha_n$&$\beta_n$\\
\hline
ChPT& 7.4 & $-2.0$ &  10.1 & $-1.2$\\ 
HBChPT &12.6& 1.3& 12.6 & 1.3\\
Empirical $E_{0+}$ &3.2 & $-0.3$ & 4.1 & $-0.4$\\
Experiment & $12.0\pm 0.6$& $1.9\mp 0.6$ & $ 12.5 \pm 1.7$& $2.7 \mp 1.8$\\ 
\hline
\end{tabular}
\end{center}
\label{tab3}
\end{table}
The result obtained by L'vov \cite{lvov93b} has been confirmed 
by the present author in
\cite{schumacher07a} by the following independent procedure. The 
photoabsorption cross-sections for the $E_{0+}$ CGLN amplitude has been
calculated  in the Born approximation and on the basis of empirical data
as shown in Figure \ref{e0plus}. Then dispersion theory has been applied
to calculate the corresponding contributions to the electric polarizabilities
of the proton and the neutron. The appropriate formulae may be found in
\cite{schumacher13}.
The results obtained in the Born approximation are 
\begin{equation}
\alpha^{E_{0+} \,{\rm Born}}_p = 7.5, \quad
\beta^{E_{0+} \, {\rm Born}}_p = -1.4,\quad
\alpha^{E_{0+}\, {\rm Born}}_n = 9.9, \quad
\beta^{E_{0+}\,{\rm Born}}_n = -1.8
\label{pionBorn}
\end{equation} 
and again are  in excellent agreement with 
the data given line 2  of Table \ref{tab3}. This means that the
equivalence of the ChPT prediction and the Born approximation
found by L'vov \cite{lvov93b} has been confirmed. By the same procedure 
the polarizabilities are also calculated from the empirical $E_{0+}$
amplitudes shown in Figure \ref{e0plus}. These results are given in 
line 4 of Table \ref{tab3}.

When comparing the predictions for the polarizabilities 
 given in Table \ref{tab3} provided by the empirical $E_{0+}$
CGLN amplitude and the corresponding Born approximation, denoted ChPT,
it is of interest to go to Figure \ref{e0plus}.
Both representations  have in common that the cross
 sections and, therefore, also the electric
polarizabilities for the neutron are larger than those for the proton.
The reason for this is that the $n\to p \pi^-$ transition leads to a larger 
electric dipole moment than the $p\to n\pi^+$ transition \cite{schumacher07a}.
Two reasons for the deviations of the empirical $E_{0+}$ amplitude 
from the Born approximation have been discussed in \cite{drechsel99}.
The first reason is that the pseudovector (PV) coupling as entering into the
Born approximation is not valid at high photon energies but has to be 
replaced by some average of PV and pseudoscalar (PS) coupling. The
second reason are $\rho$- and $\omega$-meson  exchanges which are
not taken into account in the Born approximation.  It is apparent that 
the degrees of freedom leading to the large difference between the empirical
 $E_{0+}$ CGLN amplitudes and the Born approximation are among those 
which are explicitly ``integrated out'' in Weinberg's approach. 
Furthermore, the use of the derivative PV coupling alone does not lead to the
 correct results for the cross sections and, therefore, also not for the 
polarizabilities.

The remarkable  agreement of the predictions based on the two versions of
chiral perturbations theory with the experimental data may lead to the
conclusion that these predictions essentially contain  a complete  
description of the polarizabilities and only need some
minor corrections in order to arrive at the final theoretical prediction
of the electromagnetic polarizabilities \cite{bernard91,bernard92}. 
This conclusion apparently is not correct.
In a later versions of chiral perturbation theory 
\cite{hildebrandt04,lensky12} which are considered in the next subsections in
more detail 
it was recognized that the resonant excitation of the
nucleon 
plays a dominant  r\^ole and provides a  large paramagnetic component. 
Furthermore, quantities  $\delta \alpha$ and $\delta \beta$ denoting the
electric and the magnetic  counterterms (c.t.), respectively, have been
introduced. These counterterms are also denoted  short-distance
contributions to nucleon polarizabilities \cite{hildebrandt04}.

\subsection{Chiral dynamics in low-energy Compton scattering off
  the nucleon versus dispersion theory}

For a comparison with  dispersion theory (DR) as outlined in the foregoing
and in \cite{schumacher13}
we start with the work of Hildebrandt et al. \cite{hildebrandt04}
from the following reasons. This work takes into account a complete list
components and provides numerical results for them. These components
are\\
(i) the nonresonant 
N$\pi$ component which is known to have the $E_{0+}$ CGLN amplitudes 
as the main part,\\
(ii) the short-distance or counterterm (c.t.) component which may tentatively 
be compared
with the $t$-channel component of DR,\\
(iii) the $\Delta$-pole component calculated by the small scale expansion
(SSE) method, which may be compared with the resonant
$s$-channel component of DR, and\\
(iv) the $\Delta\pi$ component which may be compared with the 
$\gamma\to \pi\Delta \to  N\pi\pi$ component of the photoabsorption cross
section.\\ 
\begin{table}[h]
\caption{Components contributing to the electromagnetic
  polarizabilities of the proton obtained in the SSE-version of HBChPT
compared with dispersion theory (DR) as outlined in the foregoing sections
and in \cite{schumacher13}. The $t$-channel components of DR
are tentatively compared with the electric, $\delta\alpha$, and magnetic,
 $\delta\beta$, counterterm (c.t.).}
\begin{center}
\begin{tabular}{|c|c|c||c|c|c|}
\hline
$\alpha_p$ &HBChPT& DR& $\beta_p$ & HBChPT & DR\\ 
\hline
N$\pi$ & +11.87 & +3.09 & N$\pi$ &  +1.25 & + 0.48\\
c.t. & -5.92 & +7.6 & c.t. & -10.68 & -7.6 \\
$\Delta$-pole & 0.0 & -0.01 & $\Delta$-pole & +11.33 & +8.56 \\
$\Delta \pi$ & +5.09 & +1.4 & $\Delta \pi$ & +0.86 & + 0.4\\  
\hline
& 11.04 &  12.08 && 2.76 & 1.84\\
\hline
\hline
\end{tabular}
\end{center}
\label{tabSSE}
\end{table}

For the N$\pi$ component of $\alpha_p$ given in line 2 of Table \ref{tabSSE}
we find a remarkable discrepancy between the results obtained by the
HBChPT-SSE   calculation and the results obtained by dispersion theory
(DR), amounting to a factor 3.8. This factor is due to the
fact that HBChPT replaces the empirical $E_{0+}$ CGLN amplitude
by a modified version
of the Born approximation
which leads to this large factor  as discussed in the foregoing
subsection. 
 
The  counterterms (c.t.) $\delta\alpha$ and $\delta\beta$
given in line 3 of Table \ref{tabSSE}
are not obtained by a parameter-free prediction, but by adjustments to
experimental data. A disadvantage of the result obtained is that the sum
$\delta\alpha+\delta\beta$  is unequal to zero 
and negative. 
Then, according to
Baldin's sum rule there should be a negative component in the total
photoabsorption cross section of the nucleon which is related to this quantity 
$\delta\alpha+\delta\beta$. Such a component does not exist.
On the other hand the predictions  in line 3 of Table \ref{tabSSE}
obtained from the $\sigma$-meson pole term of the $t$-channel (columns DR)
obey the
condition $\alpha^t+\beta^t=0$.

The paramagnetic polarizability given in  line 4 of Table \ref{tabSSE}
shows that the SSE-method leads to the right order of magnitude for this
quantity. The deviation from the prediction obtained via dispersion theory
amounts to  32\%. 

This is different for the $\Delta\pi$ component which is too big by a factor of
3.6 in case of the electric polarizability and by a factor of 2 in case of the 
magnetic polarizability. In this case similar effects may play a r\^ole 
as in case of the N$\pi$ component.

Summarizing it may be stated that the HBChPT-SSE version of chiral perturbation
theory as discussed in this sections has the advantage of containing 
all the degrees of freedom which
also are expected from the point of view of dispersion theory. There is the
nonresonant N$\pi$ component, a formal  analog  (c.t.) of the $t$-channel
component, a component from resonant excitation of the nucleon and a
$\Delta\pi$ component. 
However, the numbers obtained certainly need improvements.

\subsection{Different variants of chiral EFT}

After the first papers on the polarizabilities of the nucleon based on chiral
EFTs had been published \cite{bernard91,bernard92}, a large numbers of further
papers appeared, revealing agreement and controversies between the methods
applied by different groups. In this connection a recent topical review may be
cited \cite{griesshammer12} which covers a major fraction of the development.
For the purpose of the present work an other recent paper is of importance
 \cite{lensky12} which gives, in a concise way, 
a deeper insights into the methodology of ChPT.

One interesting piece of information is given in terms of counterterms (c.t.)
$\delta\alpha^{(p)}_{E1}$ and   $\delta\beta^{(p)}_{M1}$ entering into the
different variants of chiral EFT. These are given in line 3 of
the following Table
\ref{lensky} together with other contributions.
\begin{table}[h]
\caption{Values for the nonresonant N$\pi$ contribution, the 
counterterm (c.t.), the $\Delta$ pole and the $\Delta\pi$ contribution to the
electric $\alpha$ and magnetic $\beta$ polarizabilities  
in different variants of $\chi$EFT considered in \cite{lensky12}}
\begin{center}
\begin{tabular}{|c|c|c||c|c||c|c||c|c||c|c||c|c|}
\hline
& I $\alpha$ &I $\beta$& II $\alpha$ & II $\beta$& 
III $\alpha$ &III $\beta$& IV $\alpha$  & IV $\beta$& V $\alpha$ 
&V $\beta$ & VI $\alpha$ & VI $\beta$\\
\hline
N$\pi$& 0.0 & 0.0 &0.0&0.0& 12.6 &1.3 & 6.9 & -1.8& 12.6 & 1.3& 6.9& -1.8\\
c.t.& 10.5 & 2.7&10.6&-4.4&-2.1&1.4& 3.6 & 4.5 & -9.8 & -7.1 & -0.8 & -1.2\\
$\Delta$ pole& 0.0 & 0.0 & -0.1 & 7.1&0.0& 0.0& 0.0&0.0&-0.1& 7.1&-0.1&7.1\\
$\Delta\pi$&0.0&0.0&0.0&0.0&0.0&0.0&0.0&0.0&7.8&1.4&4.5&-1.4\\
\hline
\end{tabular}
\end{center}
\label{lensky}
\end{table}
The variants considered in \cite{lensky12}  are as follows:\\
I: The results for Compton scattering with nucleon and $\pi^0$ Born 
graphs (Tree graphs), plus
polarizabilities as given by the counterterm (c.t.).\\
II: Tree graphs plus the effects of the (dressed) $\Delta$ $s$- and
$u$-channel pole graphs.\\
III: Tree graphs plus $\pi$N loops: the $O(e^2\delta^2)$ calculation in 
heavy-baryon $\chi$EFT without  an explicit $\Delta$ degree of freedom.\\
IV: as (III) but with relativistic nucleon propagator in the $\pi$N 
and $\pi\Delta$  loops. \\
V: the $O(e^2\delta^3)$ calculation in heavy-baryon $\chi$EFT with explicit
$\Delta$, including tree graphs, $\Delta$ poles and HB $\pi$N and $\pi\Delta$
loops.\\
VI:  as (V), but with relativistic nucleon propagators in the $\pi$N and
$\pi\Delta$  loops.

According to the authors \cite{lensky12} 
all of these variants of $\chi$EFT are based on the same low-energy symmetries
of QCD and in this sense are equivalent. But  only the last two 
versions V and VI in Table \ref{lensky} are
realistic calculations which can be compared with experimental Compton
differential cross sections in the resonance region
and some way below. A comparison with experimental data at angles from
$\theta_{\rm lab}= 60^\circ$ to  $135^\circ$  shows that these  two variants 
of $\chi$EFT which  both include the leading $\pi$N loop effects and an explicit
$\Delta$ are quite similar, provided the counterterms (c.t.) are included and
adjusted to yield identical values for the scalar dipole polarizabilities.
However, the values found for the two sets of 
counterterms in the $\chi$EFT variants
considered here are rather different. They are particularly large in the 
$\pi^{\rm HB}\Delta$ calculation (column V) , especially as compared to the
relativistic $\pi\Delta$ one (column VI). 

The statements of the authors \cite{lensky12} cited in the forgoing paragraph
indicate that in case of $\chi$EFTs  a major part of the interest is directed 
to the low-energy symmetries of QCD, whereas the precision of the numbers 
attributed to the different components of the polarizabilities is of
secondary importance. This   attitude contrasts with the aim of the
non-subtracted dispersion theory where methods have been developed which lead
to a high precision of the different components of the polarizabilities. 
For the data analysis certainly high precision is required.

\section{Discussion and conclusions}

(i) The analysis of Compton scattering and polarizabilities in terms of
nonsubtracted dispersion relations is precise and well tested experimentally
in a large angular range and at energies up to 1 GeV. In this respect the 
nonsubtracted dispersion theory differs from the subtracted dispersion theory
\cite{drechsel03}  where the latter
looses validity already at the peak energy of the
$\Delta$ resonance.\\
(ii) The   polarizabilities of the nucleon
are determined from  experimental data in two ways, firstly by adjusting
the predictions of the nonsubtracted dispersion theory to the 
experimental differential cross
sections for Compton scattering in the low-energy domain below the
$P_{33}(1232)$ peak and secondly by 
calculating them from the $s$-channel and $t$-channel singularities 
as provided by
high-precision analyses of CGLN amplitudes and well known properties
of the scalar $t$-channel. In the latter case no use is made of
Compton scattering differential cross sections.
The results obtained by these two methods are in  excellent agreement with
each other.

The paper of Beane et al. \cite{beane03} cited in the Introduction 
is essentially of the variety III in Table \ref{lensky}. The amplitude for 
unpolarized Compton scattering is expanded in powers of the parameter $Q$
used in EFTs, representing a typical external momentum. 
To ${\cal O}(Q^3)$ the polarizabilities are given by pion-loop
effects and lead to the HBChPT prediction. At ${\cal O}(Q^4)$ there are new
long-range contributions to these polarizabilities. Four new parameters also
appear which encode contributions of short-distance physics to the
spin-independent polarizabilities. Thus one needs four pieces of experimental
data to fix these four short-distance contributions. After these adjustment to
experimental data have been carried out good fits to experimental differential
cross sections have been obtained.
The paper of Mc Govern et al. \cite{mcgovern13} cited in the Introduction
includes effects of the $\Delta$-resonance and, therefore, essentially 
is of the variety V in Table \ref{lensky}.

The foregoing analysis has shown that by ``integrating out'' 
$\rho$- and $\omega$- degrees of freedom and by using the derivative
coupling alone, 
the $N\pi$ component of the electric polarizability predicted by ChPT is
increased by a factor $\sim 2.4$ compared to the correct empirical value.
Furthermore, by carrying out the kinematical modifications  entering into
HBChPT a further increase by a factor  of 1.7 (proton) and 1.2 (neutron)
is obtained. The  ``integrating out'' of the $\sigma$-meson   leads to the
omission of the major part, $\alpha^t=+7.6$, of the electric polarizability
and to the omission of the diamagnetic polarizability $\beta^t=-7.6$.

\section*{Acknowledgment}

The author is indebted to A.I. L'vov, V. Pascalutsa and H.W. Griesshammer
for valuable comments.


\clearpage
\newpage
\begin{thebibliography}{99}
  

\bibitem{schumacher13}
M. Schumacher, M.D. Scadron, Fortschr.  Phys. \textbf{61} (2013) 703, 
arXiv:1301.1567
[hep-ph].  

\bibitem{hearn62}
A.C. Hearn, E. Leader, Phys. Rev. \textbf{126} (1962) 789; R. K\"oberle,
Phys. Rev. \textbf{166} (1968) 1558.


\bibitem{bernabeu74}
J. Bernabeu, T.E.O Ericson, C. Ferro Fontan, Phys. Lett. \textbf{49} (1974)  B
 381; J. Bernabeu, B. Tarrach, Phys. Lett. \textbf{69} B (1977)
 484.


\bibitem{lvov97}
A.I. L'vov, V.A. Petrun'kin, M. Schumacher, Phys. Rev. C \textbf{55} (1997)
359.


\bibitem{galler01}
G. Galler et al., Phys. Lett. B \textbf{503} (2001)  245.

\bibitem{wolf01}
S. Wolf et al., Eur. Phys. J. A \textbf{12} (2001)   231.


\bibitem{drechsel03}
D. Drechsel, B. Pasquini, M. Vanderhaeghen, Phys. Rept. \textbf{378}
(2003) 99,
arXiv:hep-ph/0212124.


\bibitem{schumacher05}
M. Schumacher, Prog.  Part.  Nucl. Phys. \textbf{55} (2005)  567,
[hep-ph/0501167]. 


\bibitem {drechsel07}
D. Drechsel, S.S. Kamalov, L. Tiator, Eur. Phys. J. \textbf{34} (2007) 69,
arXiv:0710.0306 [nucl-th].


\bibitem{beane03}
S. R. Beane, M. Malheiro, J.A. McGovern, D.R. Phillips, U. van Kolck,
Phys. Lett. B \textbf{567} (2003)  200; Erratum: Phys. Lett. B \textbf{607}
(2005)  320, arXiv:nucl-th/0209002. 

\bibitem{mcgovern13}

J.A. McGovern, D.R. Phillips, H.W. Griesshammer, Eur. Phys. J. 
A \textbf{49} (2013) 12, arXiv:1210.4104 [nucl-th].



\bibitem{PDG}
J. Beringer, et al.
(Particle Data Group) Phys. Rev. D \textbf{86} (2012)  010001 
and   partial update for the 2014 edition.



\bibitem{lvov01}
A.I. L'vov, S. Scherer, B. Pasquini, C. Unkmeir, D. Drechsel, Phys. Rev.
C \textbf{64} (2001)  015203, arXiv:hep-ph/0103172.

\bibitem{levchuk04}
M.I. Levchuk, private communication (2004).


\bibitem{schumacher07b}
M. Schumacher, Eur. Phys. J. A \textbf{34} (2007)  293, 
arXiv:0712.1417 [hep-ph].

\bibitem{schumacher09}
M. Schumacher, Nucl. Phys. A \textbf{826} (2009)  131, 
arXiv:0905.4363 [hep-ph].

 
\bibitem{schumacher06}
M. Schumacher, Eur. Phys. J. A \textbf{30} (2006) 413, Erratum
\textbf{32} (2007) 121, [hep-ph/0609040]; M.I. Levchuk, A.I. L'vov,
A.I. Milstein, M. Schumacher, Proceedings of the Workshop on the Physics of
Excited Nucleons, World Scientific, NSTAR 2005 (2005)  389, [hep-ph/0511193].


\bibitem{nambu61}
Y. Nambu, G. Jona-Lasinio, Phys. Rev. \textbf{122} (1961) 345. 

\bibitem{vogl91}
U. Vogl, W. Weise, Prog. Part. Nucl. Phys. \textbf{27} (1991) 195.

\bibitem{klevansky92}
S. P. Klevansky, Review of Modern Physics \textbf{64} (1992) 649.

\bibitem{hatsuda94}
Tetsuo Hatsuda, Teiji Kunihiro, Physics Reports \textbf{247} (1994) 221.

\bibitem{bijnens96}
Johan Bijnens, Physics Reports \textbf{265} (1996) 369.

\bibitem{gell-mann60}
M. Gell-Mann, M. Levy, Nuovo Cim. \textbf{16} (1960) 705.

\bibitem{alfaro73}
V. de Alfaro, S. Fubini, G. Furlan, C. Rosetti, in Currents in Hadron Physics
(North-Holland Publ. Amsterdam, 1973) chap.5.

\bibitem{scadron13}
M.D. Scadron, G. Rupp, R. Delbourgo, Fortschr.  Phys.
\textbf{61} (2013) 994, arXiv:1309.5041 [hep-ph].


\bibitem{weinberg79}

Steven Weinberg, Physica \textbf{96A} (1979) 327.

\bibitem{bernard91}
V. Bernard et al., Phys. Rev. Lett. \textbf{67} (1991) 1515.

\bibitem{bernard92}
V. Bernard et al., Nucl. Phys. B \textbf{373} (1992) 346.


\bibitem{lvov93b}
A.I. L'vov, Phys. Lett. B \textbf{304} (1993) 29.





\bibitem{schumacher07a}
M. Schumacher, Eur. Phys. J. A \textbf{31} (2007) 327, arXiv:0704.0200
[hep-ph]. 

\bibitem{hanstein98}
D. Hanstein, D. Drechsel, L. Tiator, Nucl. Phys. A \textbf{632} (1998) 561. 


\bibitem{drechsel99}
D. Drechsel, O. Hanstein, S.S. Kamalov, L. Tiator, Nucl. Phys. A
\textbf{645} (1999) 145.

\bibitem{hildebrandt04}
R. P. Hildebrandt, et al., Eur. Phys. J. A \textbf{20} (2004) 293,
[nucl.-th/0307070] 


\bibitem{lensky12}
V. Lensky, J. M. McGovern, D. R. Phillips, V. Pascalutsa, Phys. Rev. C
\textbf{86} (2012) 048201, arXiv:1208.4559 [nucl-th].


\bibitem{griesshammer12}
H. W. Griesshammer, J. A. McGovern, D. R. Phillips, G. Feldman, Prog. 
Part. Nucl.  Phys. \textbf{67} (2012) 841, arXiv:1203.6834 [nucl-th].







\end{thebibliography}
\end{document}